\documentclass{llncs}
\usepackage{makeidx}  % allows for indexgeneration

\usepackage{amsmath}
\usepackage{amssymb}
\usepackage{hyperref}

\begin{document}
\title{Need Polynomial Systems be Doubly-exponential?}
\titlerunning {Need Polynomial Systems be Doubly-exponential?}
% abbreviated title (for running head)

\author{James H. Davenport\inst{1} \and Matthew England\inst{2} }
\authorrunning{J.H. Davenport and M. England} 
% abbreviated author list (for running head)
\tocauthor{J.H. Davenport and M. England}
\institute{Department of Computer Science, \\
University of Bath, Bath, BA2 7AY, UK\\
\email{J.H.Davenport@bath.ac.uk},\\ 
WWW home page: \url{http://people.bath.ac.uk/masjhd/}
\vspace*{0.1in}
\and
School of Computing, Electronics \& Maths, \\ Faculty of Engineering, Environment \&
Computing, \\ Coventry University, Coventry, CV1 5FB, UK \\
\email{Matthew.England@coventry.ac.uk},\\ 
WWW home page: \url{http://computing.coventry.ac.uk/~mengland/}
}

\maketitle              % typeset the title of the contribution

\begin{abstract}
Polynomial Systems, or at least their algorithms, have the reputation of being doubly-exponential in the number of variables [Mayr and Mayer, 1982], [Davenport and Heintz, 1988]. Nevertheless, the Bezout bound tells us that that number of zeros of a zero-dimensional system is singly-exponential in the number of variables. How should this contradiction be reconciled?

We first note that [Mayr and Ritscher, 2013] shows that the doubly exponential nature of Gr\"{o}bner bases is with respect to the dimension of the ideal, not the number of variables.  This inspires us to consider what can be done for Cylindrical Algebraic Decomposition which produces a doubly-exponential number of polynomials of doubly-exponential degree. 

We review work from ISSAC 2015 which showed the number of polynomials could be restricted to doubly-exponential in the (complex) dimension using McCallum's theory of reduced projection in the presence of equational constraints.  We then discuss preliminary results showing the same for the degree of those polynomials.  The results are under primitivity assumptions whose importance we illustrate.

\end{abstract}

%%% MAX 8 PAGES

%\newpage

\section{Introduction}
\label{SEC-Intro}

We consider the title question for two of the main tools for polynomial systems: \emph{Gr\"obner Bases} (GB) and \emph{Cylindrical Algebraic Decomposition} (CAD).  For both the common claims of ``doubly exponential'', refers to ``doubly exponential in the number of variables $n$''.  All other dependencies, on polynomial degrees $d$, polynomial coefficient length $l$, or number of polynomials $m$, are themselves polynomial in these quantities (albeit with the exponent of $d$ and $m$ possibly exponential in $n$).  

In Section \ref{SEC-GB} we recall recent improvements to the analysis for GB which inspires us to revisit the complexity of CAD in Section \ref{SEC-CAD}.  Here we describe how recent work for CAD in the presence of equational constraints (equations logically implied by the input) allows for a more subtle analysis. The progress is under the assumption of primitive equational constraints and in Section \ref{SEC-Primitive} we elaborate on the importance of this.

\section{Gr\"obner Bases}
\label{SEC-GB}

A \emph{Gr\"{o}bner Basis} (GB) is a particular generating set of an ideal $I$ (defined with respect to a monomial ordering).  One definition is that the ideal generated by the leading terms of $I$ is generated by the leading terms of the GB.  GB theory allows properties of the ideal to be deduced such as dimension and number of zeros and so are one of the main practical tools for working with polynomial systems.  Introduced by Buchberger in his PhD thesis of 1965 \cite{Buchberger2006}; there has been much research to improve and optimise GB calculation, with the $F_5$ algorithm \cite{Faugere2002} perhaps the most used approach currently. 

It is common (and the authors have done this, to write) ``\cite{MM82} shows that the computation of Gr\"obner bases is doubly exponential in the number of variables''.  It is unfortunately also common simply to write ``\cite{MM82} shows that the computation of Gr\"obner bases is doubly exponential'', which while strictly correct if one counts the number of bits in a suitable encoding, is not particularly helpful.

However, we have known for a long time that the complexity of a Gr\"obner base of a zero-dimensional ideal is ``only'' singly-exponential in $n$ \cite{Lazard1983}.
These days, a much better reference is \cite{MR13}, which establishes both upper and lover bounds which are \emph{singly} exponential in $n$, but \emph{doubly} exponential in $r$, the actual dimension of the ideal. Clearly $r\le n$, and only in the worst case is $r=n$.

Though we are currently unable to capitalise on the fact, we note that the examples of  \cite{MM82,MR13} are of non-radical ideals. The effective Nullstellensatz of \cite{Kollar1988} is only singly-exponential in the number of variables for membership in the \emph{radical of} an ideal, giving us reason to believe it may be possible to prove a singly-exponential bound for radical ideals.  GB technology is also needed to realise similar improvements to the complexity bound of CAD, as discussed next.

%\newpage

\section{Cylindrical Algebraic Decomposition}
\label{SEC-CAD}
% Basically a summary of ISSAC 2015 and CASC

\subsection{Background}

A \emph{cylindrical algebraic decomposition} (CAD) is a \emph{decomposition} of $\mathbb{R}^n$   into cells.  The cells are arranged \emph{cylindrically}, meaning the projections of any pair with respect to the given ordering are either equal or disjoint.  We assume variables labelled according to their ordering (so the projections considered are $(x_1,\ldots,x_{\ell})\rightarrow(x_1,\ldots,x_k)$ for $k<\ell$) with the highest ordered variable present said to be the \emph{main variable}.  Finally, by \emph{algebraic} we mean semi-algebraic: each cell can be described with a finite sequence of polynomial constraints.  

A CAD is produced to be invariant for input; originally \emph{sign-invariant} for a set of input polynomials (so on each cell each polynomial is positive, zero or negative), and more recently \emph{truth-invariant} for input Boolean-valued  formulae built from the polynomials (so on each cell each formula is either true or false).  Unlike Gr\"{o}bner Bases we may now consider general polynomial systems instead of just equations.

CAD usually involves two phases.  The first {\em projection}, applies operators recursively on  polynomials, each time producing a set with one less variable which together define the {\em projection polynomials}.  These are used in the second phase, {\em lifting}, to build CADs incrementally by dimension.  First a CAD of the real line is built according to the real roots of the univariate polynomials. % (those in $x_1$ only).  
Next, a CAD of $\mathbb{R}^2$ is built by repeating the process over each cell in $\mathbb{R}^1$ with the bivariate polynomials %in ($x_1,x_2)$ 
evaluated at a sample point of the cell in $\mathbb{R}^1$.  We call the cells where a polynomial vanishes {\em sections} and those regions in-between {\em sectors}, which together form the {\em stack} over the cell.  Taking the union of these stacks gives the CAD of $\mathbb{R}^2$.  The process is repeated until a CAD of $\mathbb{R}^n$ is produced.  In each lift we extrapolate the conclusions drawn from working at a sample point to the whole cell requiring validity theorems for the projection operator used.

CAD was originally introduced by Collins for quantifier elimination (QE) in real closed fields \cite{ACM84I} with applications since ranging from parametric optimisation \cite{FPM05} and epidemic modelling \cite{BENW06}, to reasoning with multi-valued functions \cite{DBEW12} and the derivation of optimal numerical schemes \cite{EH14}. 
There has been much work on improving Collins' original approach most notably refinements to the projection operator \cite{McCallum1998} \cite{Brown2001a}, \cite{HDX14}; early termination of lifting \cite{CH91} \cite{WBDE14}; and symbolic-numeric schemes \cite{Strzebonski2006}, \cite{IYAY09}.  Some recent advances include dealing with multiple formulae \cite{BDEMW13}, \cite{BDEMW16}; local projection \cite{Brown2013}, \cite{Strzebonski2014a}; decompositions via complex space \cite{CMXY09}, \cite{BCDEMW14}; and the development of heuristics for CAD problem formulation \cite{BDEW13}, \cite{EBCDMW14}, \cite{WEBD14} including machine learned approaches \cite{HEWDPB14}.

\subsection{Complexity}

CAD has long been known to have worst case complexity doubly exponential \cite{BD07},\cite{DH88}. Suppose the input consists of $m$ polynomials (perhaps derived from formulae) in $n$ variables of maximum degree $d$ in any one variable.  Section 2.3 of \cite{BDEMW16} describes in detail how the complexity of CAD algorithms may be measured in terms of a bound on the total number of cells produced (closely correlated to the timings but allowing for simpler implementation independent comparisons) based on improvements to techniques introduced by McCallum's thesis.  In particular, the dominant term in that bound for a sign-invariant CAD produced using the algorithm of \cite{McCallum1998} is
\begin{equation}
\label{eq:BoundSI}
(2d)^{2^{n}-1}m^{2^{n}-1}2^{2^{n-1}-1}.
\end{equation}
I.e. the CAD grows doubly exponentially with the number of variables $n$.
The analysis shows that by the end of the projection stage we have $M$ polynomials in $\mathbb{R}^1$, each of degree $D$, where $D=d^{2^{O(n)}}$ and $M=m^{2^{O(n)}}$. 
However, \cite{DH88} \cite{BD07} respectively find lower bounds with $D=d^{2^{\Omega(n)}}$ and $M=m^{2^{\Omega(n)}}$ with the underlying polynomials all simple, showing that the doubly-exponential difficulty of CAD resides in \emph{the complicated number of ways simple polynomials can interact}. 

\subsubsection*{So need CAD be doubly exponential?}  Given the previous discussion the answer is yes, but as with GB we need not settle for ``doubly exponential \emph{in the number of variables} $n$''.  We might hope for ``doubly exponential \emph{in the dimension}'', but this is thwarted by the fact that the examples of  \cite{BD07},\cite{DH88} are in fact zero-dimensional.  Nevertheless, we can take advantage of certain dimensional reductions when made explicit through the identification of \emph{equational constraints} (ECs), polynomial equations logically implied by formulae.

The presence of an EC restricts the dimension of the solution space and so we may expect the CAD to be doubly exponential in $n-\ell$ where $\ell$ is the number of ECs taken advantage of.  
Of course, we would no longer be building CADs sign-invariant for polynomials but ones truth-invariant for formulae.   The present authors have demonstrated this first for the part of the bound dependent on $m$ (number of polynomials) in \cite{EBD15} and then for the part dependent on $d$ (maximum degree) in \cite{ED16} (work currently submitted for publication).

\subsection{CAD with multiple ECs}

Collins noticed that in the presence of an EC a truth-invariant CAD need only be sign-invariant throughout for the defining polynomial of the EC with other polynomials sign-invariant only on the sections of that polynomial \cite{Collins1998}. This led McCallum to develop restrictions to his projection operator from \cite{McCallum1998} in \cite{McCallum1999a} (for the first projection) and \cite{McCallum2001} (for subsequent projections).   See \cite[Section 2.1]{EBD15} for a more detailed summary of this theory.  These operators work with a single EC and so the CAD algorithm may take advantage of only one in each main variable.  
However, \cite{McCallum2001} also introduced a process to derive ECs in lower main variables based on the observation that the resultant of the polynomials defining two ECs itself defines an EC.  

In \cite{EBD15} the present authors reviewed the theory of reduced projection operators.  In particular we introduced two refinements to the lifting phase of CAD which follow from  McCallum's theory of reduced projection operators:
\begin{enumerate}
\item Minimising lifting polynomials: When lifting \emph{to} $\mathbb{R}^k$ if there exists an EC with main variable $k$ then we need only lift with respect to (isolate roots of) this.  
\item Minimising real root isolation: When lifting \emph{over} $\mathbb{R}^k$ if there exists an EC with main variable $k$ then we need only isolate real roots over sections (allowing sectors to be trivially lifted to a cylinder).
\end{enumerate}
These refinements require us to discard two embedded principles of CAD:
\begin{itemize}
\item That the projection polynomials are a fixed set: we now differ the polynomials used in projection from lifting and keep track of which relate to ECs. 
\item That the invariance structure of the final CAD can be expressed in terms of sign-invariance of polynomials:  The final CAD may not no longer be sign-invariant for any one polynomial polynomials, even ECs, but is still guaranteed to be truth invariant for the formula.  
\end{itemize}

In \cite[Section 5]{EBD15} we used the complexity analysis techniques of \cite{BDEMW16} to show that a CAD in which the first $\ell$ projections had a designated EC had dominant term complexity bound of the form $(2d)^{\mathcal{O}(2^{n})}(2m)^{\mathcal{O}(2^{n-\ell})}$.  I.e. we have reduced the number of polynomials involved accordingly but not their degree.

The present authors considered what could be done with respect to the degree recently in \cite{ED16}.  The theory of iterated resultants as considered by Bus\'{e} and Mourrain \cite{BM09} suggested that the iterated univariate resultants produced by CAD (and in particular in the identification of ECs for subsequent projections) were more complicated that the information they needed to encode.  The true multivariate resultants were contained as a factor and grow in degree exponentially rather than doubly exponentially.  The key result had to be adapted from \cite{BM09} to change the arguments from total degree in all variables to the degree in at most one variable required for bounding the number of CAD cells produced.

The authors proposed using GB technology for the generation of the ECs in subsequent projections to realise this limit in degree growth.  This leads to the other projection polynomials growing exponentially in $O(\ell^2)$ but remember that these are not used during lifting (\cite{EBD15} improvement (1) from above) and thus not counted towards the cell count bounds (although they do boost the degree of polynomials involved in projections without ECs).  The outcome of this approach is a dominant term complexity bound of the form 
$(\ell d)^{\mathcal{O}(2^{n-\ell})}(2m)^{\mathcal{O}(2^{n-\ell})}$.

\subsubsection*{Restrictions}  There are some restrictions to the work as acknowledged in \cite{EBD15} and \cite{ED16}.  First, the analysis assumes the designated ECs are in strict succession at the start of projection.  This restrictions was made to ease the complexity analysis (with the formal algorithm specification and implementations not adhering).  

The substantial  restriction is that the theory of CAD with multiple ECs is only developed for primitive ECs.  Possibilities to remove this restriction are discussed in \cite{EBD15} and could involve leveraging the TTICAD theory of \cite{BDEMW13} \cite{BDEMW16}.  A TTICAD (truth-table invariant CAD) allows for savings from ECs when building a CAD for multiple formulae at once.  Currently the theory is only developed for ECs in the main variable of the system and so an analogous extension to subsequent projections is first required for TTICAD itself.

% JHD: it's 00:23 and I'm no longer sure of this part. I'll go to bed and think!

\section{The primitivity restriction}
\label{SEC-Primitive}

We finish by considering the classic complexity results of \cite{BD07},\cite{DH88} in light of the above recent progress.  We see the importance of the aforementioned primitivity restriction.

The examples in both \cite{BD07} and \cite{DH88} rest on the following construction.
Let $P_k(x_k,y_k)$ be the statement $x_k=f(y_k)$ and then define recursively
\begin{align}
&P_{k-1}(x_{k-1},y_{k-1}) := \label{eq:H} \\
&\begin{array}{c}
\cr\underbrace{\exists z_k\forall x_k\forall y_k}_{Q_k}\underbrace{\left((y_{k-1}=y_k\land x_{k}=z_k)\lor(y_{k}=z_k\land x_{k-1}=x_k)\right)}_{L_k}\Rightarrow P_k(x_k,y_k).
\end{array} \nonumber
\end{align}
%\begin{equation}
%\label{eq:H}
%\begin{array}{c}
%P_{k-1}(x_{k-1},y_{k-1}):=\cr\underbrace{\exists z_k\forall x_k\forall y_k}_{Q_k}\underbrace{\left((y_{k-1}=y_k\land x_{k-1}=z_k)\lor(y_{k-1}=z_k\land x_{k-1}=x_k)\right)}_{L_k}\Rightarrow P_k(x_k,y_k).
%\end{array}
%\end{equation}
This is $\exists z_k \left(z_k=f(y_{k-1})\land x_{k-1}=f(z_k)\right)$, i.e. $x_{k-1}=f(f(y_{k-1}))$. It is repeated nesting of this procedure that builds the doubly-exponential growth, so that 
\begin{equation}\label{eq:H2}
Q_{k-1}L_{k-1}\Rightarrow \left(Q_kL_k\Rightarrow P_k(x_k,y_k)\right),
\end{equation}
gives $x_{k-2}=f(f(f(f(y_{k-2}))))$ etc.
Rewriting (\ref{eq:H2}) in prenex form gives 
\begin{equation}\label{eq:H3}
Q_{k-1}Q_k \neg L_{k-1}\lor\neg L_k\lor P_k(x_k,y_k).
\end{equation}
The negation of  (\ref{eq:H3}) is therefore
\begin{equation}\label{eq:H4}
\overline Q_{k-1}\overline Q_k  L_{k-1}\land L_k\land\neg P_k(x_k,y_k),
\end{equation}
where the $\overline{\strut\quad}$ operator interchanges $\forall$ and $\exists$.

Now, $L_k$ can be rewritten as
\begin{align}
\label{eq:H5}
L_k &= (y_{k-1} = y_k\lor y_{k}=z_k) \land (y_{k-1}=y_k\lor x_{k-1}=x_k) \nonumber \\
&\qquad \land (x_{k}=z_k\lor y_{k}=z_k)\land(x_{k}=z_k\lor x_{k-1}=x_k)
\end{align}
and further
\begin{align}
\label{eq:H6}
L_k &= (y_{k-1}-y_k)( y_{k}-z_k)=0\land(y_{k-1}-y_k)( x_{k-1}-x_k)=0 \nonumber \\
&\qquad \land (x_{k}-z_k)( y_{k}-z_k)=0\land(x_{k}-z_k)( x_{k-1}-x_k)=0,
\end{align}
which shows $L_k$ to be a conjunction of (imprimitive) equational constraints. This is true for any $L_i$, hence  the propositional part of (\ref{eq:H4}) is a conjunction of eight equalities, mostly imprimitive, and $\neg P_k(x_k,y_k)$.  Furthermore there are equalities whose main variables are the first variables to be projected if we try to produce a quantifier-free form of (\ref{eq:H4}). But the quantifier-free form of (\ref{eq:H4}) describes the complement of the semi-algebraic varieties in \cite{BD07} or \cite{DH88}  (depending which $P_k$ we take) and these have doubly-exponential complexity in $n$.

The discussion of this section shows the relevance of the primitivity restriction discussed at the end of the previous section and imposed in the work of \cite{EBD15}, \cite{ED16}.  It may be more than a technicality to remove them.

\subsubsection*{Acknowledgements}

This work was originally supported by EPSRC grant: EP/J003247/1 and is now supported by EU H2020-FETOPEN-2016-2017-CSA project $\mathcal{SC}^2$ (712689).  
We are also grateful to Professor Buchberger for reminding JHD that Gr\"obner Bases were applicable to CAD complexity.

\bibliographystyle{plain}

\end{document}